\documentclass[12pt]{article}
\usepackage{amssymb}
\usepackage{graphicx}
\usepackage{graphics}
\usepackage{cite}
\usepackage{ulem}
\usepackage{amsmath}
\usepackage{mathtools}
\usepackage{color}

\def\d{\partial}
\def\l{\left(}                                    
\def\r{\right)}

\newcommand{\be}{\begin{equation}}
\newcommand{\ee}{\end{equation}}
\newcommand{\ba}{\begin{align}}
\newcommand{\ea}{\end{align}}
\newcommand{\bg}{\begin{gather}}
\newcommand{\eg}{\end{gather}}
\newcommand{\bseq}{\begin{subequations}}
\newcommand{\eseq}{\end{subequations}}

\renewcommand{\tanh}{\mathop{\rm th}\nolimits}

\begin{document}
\title{On the dark radiation problem in the axiverse}
\author{Dmitry Gorbunov$^{1,2}$, Anna Tokareva$^{1,3}$\\
\mbox{}$^{1}${\small Institute for Nuclear Research of Russian Academy of
  Sciences, 117312 Moscow,
  Russia}\\  
\mbox{}$^{2}${\small Moscow Institute of Physics and Technology, 
141700 Dolgoprudny, Russia}\\ 
\mbox{}$^{3}${\small Ecole Polytechnique Fб╢edб╢erale de Lausanne, CH-1015, Lausanne, Switzerland}\\ 
}

\maketitle

\begin{abstract} 
String scenarios generically predict that we live in a so called
axiverse: the Universe with about a hundred of light axion species
which are decoupled from the Standard Model particles. However, the
axions can couple to the inflaton which leads to their production
after inflation. Then, these axions remain in the expanding Universe
contributing to the dark radiation component, which is severely
bounded from present cosmological data.  We place a general constraint
on the axion production rate and apply it to several variants of
reasonable inflaton-to-axion couplings.  The limit merely constrains
the number of ultralight axions and the relative strength of
inflaton-to-axion coupling. It is valid in both large and small field
inflationary models irrespectively of the axion energy scales and
masses. Thus, the limit is complementary to those associated with the
Universe overclosure and axion isocurvature fluctuations.  In
particular, a hundred of axions is forbidden if inflaton universally
couples to all the fields at reheating. In the case of gravitational
sector being responsible for the reheating of the Universe (which is a
natural option in all inflationary models with modified gravity), the
axion production can be efficient. We find that in the Starobinsky
$R^2$-inflation even a single axion (e.g. the standard QCD-axion) is
in tension with the Planck data, making the model inconsistent with
the axiverse. The general conclusion is that an inflation with
inefficient reheating mechanism and low reheating temperature may be in
tension with the presence of light scalars.

\end{abstract}
\section{Introduction}
In order to solve several problems of the hot Big Bang cosmology, the
inflationary stage of the Universe evolution has been
proposed\,\cite{Starobinsky,Guth}. The simplest way to organize the
close-to-exponential expansion of the Universe is to exploit the
scalar field (inflaton) slowly rolling towards the minimum of the
potential\,\cite{Linde,Albrecht}.  Recent data from the Planck
satellite put such strong constraints on the inflaton potential that
the simplest quartic and quadratic cases turn out to be excluded
\cite{Ade:2015lrj}. The best agreement with the data is still
exhibited by the large field inflation provided by the exponentially
flat potential. Such a potential naturally arises in different models
with the modified gravitational sector
\cite{Starobinsky,Spokoiny:1984bd,Bezrukov:2007ep} as well as in the
supergravity framework\,\cite{Ferrara:2010yw, Ellis:2013xoa,
Ketov:2010qz}.  All these models, being non-renormalizable, require an
ultraviolet completion which is often thought to be a string theory.

However, the common prediction of many string scenarios is the
existence of a plenty of light scalars (axions) arising as zero modes
of antisymmetric gauge fields on the compactified dimensions
\cite{Svrcek:2006yi}. They inherit a perturbative shift symmetry
violated by instanton contributions
\cite{Dine:1986zy,Kallosh:1995hi}. One of these scalars can play a
role of QCD axion explaining the zero value of the QCD
$\theta$-angle. The number of light axions is determined by the
topology of extra dimensions and is likely to be about hundred\,\cite{Arvanitaki:2009fg}.

Many potentially observable consequences of the string axiverse were
discussed in literature \cite{Arvanitaki:2009fg,Acharya:2015zfk}. Here
we explore whether many light scalars are compatible with
inflationary models. String axions can contribute to dark matter
component, see e.g.\,\cite{Acharya:2010zx}. In order to avoid
overproduction and suppress isocurvature fluctuations, one constrains
the axion masses, couplings and initial conditions at the inflationary
stage.  At the same time, light scalars produced at reheating can
contribute to the dark radiation, which implies an additional bound on
the model parameters. Dark radiation amount at Big Bang
Nucleosynthesis (BBN) and recombination is now strictly bounded by the
Planck data \cite{Ade:2015xua}. In generic string scenarios, the
number of effective relativistic degrees of freedom at these epochs is
predicted to be much larger than the Planck results \cite{Ade:2015xua}
allow, the discrepancy is mostly due to the contribution of moduli
decays \cite{Acharya:2015zfk}. One can suppose, however, that the
supersymmetry breaking scale, as well as the masses of all moduli, are
larger than the Hubble parameter at inflation. In this case, the
moduli are not produced and the axions are created only by the
inflaton decay. We show that in this case, if the inflaton couples to
SM particles only through the gravity (i.e. via suppressed by Planck
mass operators, which is natural, e.g. for $F(R)$ inflationary
models), then even one light scalar is in tension with the recent
Planck data. Thus, in the axiverse the inflaton must couple to matter
much stronger. In that case, we put a lower bound on the reheating
temperature which makes the existence of hundreds of light scalars
consistent with the Planck data.

\section{Reheating in the axiverse}
\label{Sec:reh}

In each model, the inflationary stage must be followed by the reheating
process eventually populating the Universe with the hot plasma of SM
particles. Therefore, the energy of the inflaton field must be somehow
transformed into the usual matter. In general, during this process
axion-like particles can well be produced. Let the 
production rate of each axion be $\Gamma_{\text{a}}$ while that of the 
SM particles be $\Gamma_{\text{SM}}$. Then, the number of additional degrees
of freedom at BBN and recombination written traditionally as the
number of additional neutrino components is
\be
\label{N_eff}
\Delta N_{\text{eff}}=N
\frac{g_{\text{reh}}}{g_{\nu}}\l\frac{g_{\text{BBN}}}{g_{\text{reh}}}\r^{4/3}
\frac{\Gamma_{\text{a}}}{\Gamma_{\text{SM}}}\,.
\ee
Here $N$ is the number of axion species, 
$g_{\nu}=2\cdot 7/8$, $g_{\text{reh}}$ and $g_{\text{BBN}}$ are the 
effective number of
relativistic degrees of freedom at reheating and nucleosynthesis
stages, respectively. Their SM values are $g_{\text{reh}}=106.75$ and
$g_{\text{BBN}}=10.75$. According to the latest Planck data, within the
concordance $\Lambda$CDM cosmological model the effective
number of relativistic species $N_{\text{eff}}$ is bounded as\,\cite{Ade:2015xua}
\be
\label{bound}
N_{\text{eff}}= 3.15\pm 0.23.
\ee 
This parameter is expected to be measured in future 
CMB polarization experiments with much higher accuracy
\cite{Abazajian:2013oma,Errard:2015cxa}. 

One can observe from \eqref{N_eff} that {\it if
the reheating is due to some universal mechanism} (i.e. inflaton decays
to all scalars including the SM Higgs with the comparable rates
$\Gamma_{\text{a}}\sim \Gamma_{\text{SM}}/4$) 
then $\Delta N_{\text{eff}}\sim 0.7 \,N$ which
{\it is clearly incompatible with} bound \eqref{bound} in the 
{\it axiverse}. 
This is the main finding of the present paper. 

The bound \eqref{N_eff}, \eqref{bound} constrains axion coupling to
inflaton and the number of axions ultrarelativistic at BBN and
recombination. It is applicable to both large- and small-field
inflation with any mass pattern in the axion sector, as far as the axions remain
ultrarelativistic. Thus, the obtained constraint is complementary to
another bounds on the axion energy scales, masses and on the energy scale of
inflation. These bounds are inferred from the Universe overclosure argument and absence
of the axion isocurvature
fluctuations\,\cite{Hertzberg:2008wr,Acharya:2010zx}.

Formula \eqref{N_eff} is exact if the axion branching ratio
$\Gamma_{\text{a}}/\Gamma_{\text{SM}}$ is constant in time at the
reheating epoch. Otherwise it is corrected by a numerical factor
of order one, which is the case if the inflaton couplings to the axion and
SM-particles are of the different nature, e.g. provided by the operators of
different dimensions. Then $\Gamma_{\text{SM}}$ in eq.\,\eqref{N_eff}
defines the age of the Universe at reheating, $t_U\sim
1/\Gamma_{\text{SM}}$ and $\Gamma_{\text{a}}/\Gamma_{\text{SM}}\sim
\Gamma_{\text{a}}t_U\ll 1 $ refers to the fraction of inflaton energy
transfered to axions. 

The case when moduli decay is responsible for the
reheating provides an example of the universal reheating mechanisms, which 
has been studied in Ref.\,\cite{Acharya:2015zfk}. Below 
we consider several other examples of models  with universal
reheating mechanisms relevant to the problem with axiverse.

\subsection{The Starobinsky model}
\label{sec:star}
The Starobinsky model of inflation historically was the first
successful model suggested for the exponential stage of the Universe
expansion \cite{Starobinsky}. However, it still provides the
predictions which are in a perfect agreement with the present
cosmological data
\cite{Ade:2015xua}. In a Jordan frame, the model is described by the
following action:
\begin{equation}
\label{original-R2}
S=-\frac{M_{\text{P}}^2}{2}\int \!\! \sqrt{-g}\;d^4x\, \left[
R-\frac{R^2}{6\,m^2} \right] + S_{\text{matter}}\;. 
\end{equation}
Here the reduced Planck mass is $M_{\text{P}}=M_{\text{Pl}}/\sqrt{8\pi}=2.4\times
10^{18}$\,GeV and $S_{\text{matter}}$ denotes the action for all matter
fields in the theory. The value of $m$, which is actually
mass of an additional scalar degree of freedom (scalaron)
responsible for inflation, is determined by the amplitude of scalar
perturbations as $m=1.3\times 10^{-5}\, M_{\text{P}}$ \,\cite{scalaron}.

After the Weyl transformation of the metric to the Einstein frame, 
\begin{equation}
\label{Weyl-transformation}
g_{\mu\nu}\rightarrow e^{\sqrt{2/3}\,\phi/M_{\text{P}}}g_{\mu\nu}\,, 
\end{equation}
action \eqref{original-R2} takes the form \cite{magnano} 
\be
S=\int{\!\!\sqrt{-g}\; d^4 x \left[ -\frac{M_{\text{P}}^2}{2}R +
\frac{1}{2}\d_{\mu} \phi \d^{\mu} \phi - V(\phi)\right]} +
\tilde{S}_{\text{matter}}\;, \ee \be V(\phi)=\frac{3 m^2 M_{\text{P}}^2}{4} \l
1-e^{-\sqrt{2/3}\phi/M_{\text{P}}}\r^2\;.  
\ee 
Here $\tilde{S}_{\text{matter}}$
is the Weyl transformed action of the matter fields. Thus, conformal
non-invariance in the matter sector naturally implies the interaction
between scalaron $\phi$ and all other particles. Fermions and
vector fields are Weyl invariant at the tree level in the high energy
limit.  Therefore, a key role in the process of reheating is
played by the light scalars: SM Higgs boson and axions in the
discussed framework (see Ref.\,\cite{Gorbunov:2013dqa} for dilaton).

Axions $a_i$, $i=1,\dots,N$,
unlike the Higgs, due to the perturbative shift
symmetry \cite{Dine:1986zy} can not be non-minimally coupled to the gravity via terms
$R a_i^2$. Thus, their kinetic terms are canonical in the Jourdan
frame\,\eqref{original-R2}, yielding the universal couplings to the
inflaton. The decay rates of 
scalaron to Higgs pair and to axion pair are\,\cite{vilenkin, Gorbunov:2012ij}
\be
\label{rates}
\Gamma_{\text{SM}}=\frac{m^3(1+6\xi_h)^2}{48\pi\,M_{\text{P}}^2}\,,~~\Gamma_{\text{a}}=
\frac{m^3}{192\pi M_{\text{P}}^2}\,,
\ee
respectively, with the SM Higgs boson possibly non-minimally coupled
to gravity via lagrangian term ${\cal L}=\xi_h R h^2/2$. One
  observes  
that the decay rates to the axions and to the SM particles (Higgs bosons) 
are comparable.

The ultrarelativistic axions produced at reheating remain in the
late Universe contributing to the energy density and pressure at BBN
and recombination as additional
\be
\label{starobinsky}
\Delta N_{\text{eff}}= 0.7 N,
\ee
neutrino flavors (we put \eqref{rates} with $\xi_h=0$ into
\eqref{N_eff}). Hence, {\it even
one additional light scalar} (for
example, standard QCD axion) is already {\it in tension with the Planck
bound} \eqref{bound} in the Starobinsky model.

To resolve the situation one can add some new $N_s$ scalars to the
matter content of the SM. Then the r.h.s. of \eqref{starobinsky} gets
suppressed by a factor $4/(4+N_s)$, which makes the Starobinsky model
in the axiverse populated by $N\sim100$ axions  consistent with the
present cosmological bounds \eqref{bound} if $N_s\sim 200$ scalars are
added.

\subsection{Inflation with non-minimal kinetic terms}

Another way to obtain the favored by the Planck results
exponentially flat potentials in a natural
way is connected with the modification of the kinetic term of the
inflaton. This idea is widely discussed in the context of supergravity
\cite{Ferrara:2010yw, Kawasaki:2000yn, Ellis:2013xoa, Ellis:2013nxa}
where non-trivial kinetic terms come from the Kahler
potential. In particular, it was realised as $\alpha$-attractors in
Refs.\,\cite{Kallosh:2013tua, Kallosh:2013yoa}, 
where the inflationary region corresponds to the
pole in the kinetic term of the inflaton. In all these models the inflaton
may be canonically normalized upon an appropriate field
transformation. In general, one can expect that the
kinetic terms of additional scalars (axions) are also
non-minimal. Thus, the action consistent with the shift symmetry of
axions $a_i$ reads 
\be
\label{axion-action}
S=\int d^4x \sqrt{-g}\l-\frac{M_{\text{P}}^2}{2} R +
\frac{(\d_{\mu}\phi)^2}{2}+\sum\limits_{i=1}^{N} f_i(\phi)
\frac{(\d_{\mu}a_i)^2}{2}-V(\phi) \r\,,
\ee
where $f_i(\phi)$ are some functions of the inflaton field.

Similarly one may expect non-renormalizable couplings to the SM fields
yielding the most relevant for reheating two-body decays of inflaton,
\be
\label{S_int}
S_{int}=\int d^4x \sqrt{-g}\l y(\phi)|D_{\mu}{\cal H}|^2-\frac{1}{4}g_j(\phi)F_{\mu\nu, j}F^{\mu\nu}_j+z_i(\phi)\bar{\psi_i}\gamma^{\mu}D_{\mu}\psi_i \r.
\ee 
Terms in the last set in \eqref{S_int} 
are proportional to the fermion masses through the equation
of motion and hence their role in the reheating is negligible.

Near the minimum of 
$V(\phi)$ (we take it to be zero) one may anticipate the expansions
\be
\begin{split}
\label{exp}
f_i(\phi)=&1+\beta_i\frac{\phi}{\Lambda}+\gamma_i\frac{\phi^2}{\Lambda^2}+\dots, \\
g_i(\phi)=&1+\delta_i \frac{\phi}{\Lambda}+\dots,~~y(\phi)=1+\gamma\frac{\phi}{\Lambda}+\dots.
\end{split}
\ee
From the point of view of the effective theory considered after
inflation such terms suppressed by cutoff scale $\Lambda<M_P$ are
naturally expected with $\beta$,$\gamma$, $\delta\sim 1$ from quantum
corrections since the potential of the inflaton (as well as the
gravity) is non-renormalizable. However, during inflation
$f_i(\phi)\approx 1$ because of the approximate shift symmetry of the
inflaton field  with Planck-favored plateau-like potential 
 providing no significant physical effects. But after
inflation the expansions \eqref{exp} start to work leading to the inflaton
decay. For example, with $\gamma\neq0$ the
reheating process can go through the decay of the inflaton to the Higgs bosons. If the
coupling to gauge bosons is suppressed for some reason we are left
with the similar case as in the Starobinsky model where the dark
radiation production is highly efficient \eqref{starobinsky}. This
model is cosmologically forbidden.

If the inflaton couples to the kinetic terms of all the matter fields
in the model with $\gamma,\,\delta_i\sim 1$, then
all the bosons will be produced roughly at equal amounts. In this
case one can
get an estimate for the axion contribution to the effective
relativistic degrees $\Delta N_{\text{eff}}$ by putting
$\Gamma_{\text{a}}/\Gamma_{\text{SM}}\sim 1/30$ in
\eqref{N_eff}. In this way one obtains $\Delta N_{\text{eff}}\sim
N/10$, which is certainly cosmologically forbidden for $N\sim 100$
given the constraint \eqref{bound}.

The problem of axion overproduction may be
avoided if the inflaton couples to the SM particles stronger than
to the axions. 
The former may be parametrized by means of the time of reheating (as we
discuss at the beginning of Sec.\,\ref{Sec:reh}) or the reheating
temperature 
$T_{reh}$, i.e. the temperature of the SM plasma at the moment when a
half of the total energy density is already in the form of radiation.

Then, if the
inflaton mass is $m$, one obtains from Eqs.\,\eqref{axion-action},\,\eqref{exp},
\be
\label{reh}
\Gamma_{\text{SM}}\simeq 3
\frac{T^2_{\text{reh}}}{\sqrt{g_{\text{reh}}} M_{\text{P}}}\,, \quad
\Gamma_{\text{a}}= \frac{\beta^2 m^3}{128\pi \Lambda^2}\,.
\ee
Substituting
\eqref{reh} into the equation \eqref{N_eff} we obtain for the amount of
dark radiation:
\be
\Delta N_{\text{eff}}= 0.024\, N\, \frac{\beta^2 m^3
  M_{\text{P}}}{\Lambda^2 T_{\text{reh}}^2}.
\ee
For the reheating temperature high enough one can see that $N\sim 100$
may still be allowed by the Planck constraints. On the contrary,
inefficient reheating with low $T_{\text{reh}}$ can easily throw the model
out of the viable range\,\eqref{bound}.

Note in passing that
the terms of the first order in the expansion \eqref{exp} may
be forbidden due to some symmetry (the simplest one is $\mathbb{Z}_2$,  
$\phi\rightarrow -\phi$). In this case, the production of axions would be
inefficient if the Universe is reheated due to the other inflaton
couplings to matter provided by a lower order operators.

We discuss this case in more details in the
next Section using the inflaton non-minimally coupled to gravity as
a realistic example.

\subsection{Inflation driven by the scalar field non-minimally coupled
  to gravity}

Although inflation models with reasonable (e.g. renormalizable without gravity)
power-law potentials are disfavoured by the Planck data, 
switching on the non-minimal
coupling of the inflaton to gravity can provide with the flat
potential suppressing the tensor modes. Models of such type are widely
discussed in the literature
(see e.g., \cite{Tsujikawa:2004my,Bezrukov:2013fca,Kallosh:2013tua})  
and include the Higgs inflation
\cite{Bezrukov:2007ep}. The action for the inflaton field $\phi$ reads
(here we neglect the possible mass term for the inflaton at large
field values),
\begin{equation}
\label{HI}
    S =\int d^4x \sqrt{-g} \l- \frac{M_{\text{P}}^2+\xi \phi^2}{2}R
    + \frac{(\partial^\mu \phi)^2}{2}-\frac{\lambda \phi^4}{4} \r. 
 \end{equation}
One can get rid of the non-minimal coupling by making use of
the metric redefinition
\begin{equation}
\label{Weyl}
   \hat{g}_{\mu\nu} = \Omega^2 g_{\mu\nu}
  \;,\quad
  \Omega^2 = 1 + \frac{\xi \phi^2}{M_{\text{P}}^2}
  \;.
\end{equation}
After that, the action in the Einstein frame takes the form
\begin{equation}
      S_{\text{E}} =\int d^4x\sqrt{-\hat{g}} \Bigg\{
    - \frac{M_{\text{P}}^2}{2}\hat{R}
    + \frac{ (\partial^\mu \chi)^2}{2}
    - U(\chi)
    \Bigg\}
    \;,
\end{equation}
where canonically normalized field $\chi$ is defined by 
\begin{equation}
\label{change}
\frac{d\chi}{d \phi}=\sqrt{\frac{\Omega^2+6\xi^2\phi^2/M_{\text{P}}^2}{\Omega^4}},
\;\;\;\;\;\text{and}\;\;\;\quad U(\chi) =
  \frac{1}{\Omega^4(\chi)}\frac{\lambda}{4}\,\phi^4(\chi)
  \;.
\end{equation}

The kinetic term of axion gets coupled to the inflaton in the Einstein frame:
\be 
\label{Laa}
L_{\text{a}}= \frac{1}{2}\Omega^2 (\d_{\mu} a)^2 = \frac{1}{2}\l
1+\frac{\xi \phi^2}{M_{\text{P}}^2}\r (\d_{\mu} a)^2.  \ee At first
sight, this coupling is quadratic in inflaton and seems to be strongly
suppressed by squared Planck scale. However, at and some time after
inflationary epoch the inflaton field takes large values, $\phi\sim
M_{\text{P}}$. This makes the second term in parenthesis in
\eqref{Laa} important to the extent which depends on the value of
nonminimal coupling $\xi$.  Taking into account this fact we study two
different cases for the value of $\xi$ which finally yield different
results.

{\bf Large non-minimal coupling, $\xi\gg 1$.} 
This case includes the model of
 Higgs inflation \cite{Bezrukov:2007ep}. In this limit for large field
 values of $\phi>M_{\text{P}}/\sqrt{\xi}$ one obtains from \eqref{change}
 \begin{equation}
 \label{stage1}
  \phi\simeq
  \frac{M_{\text{P}}}{\sqrt{\xi}}\exp\left(\frac{\chi}{\sqrt{6}M_{\text{P}}}
\right), \quad U(\chi) = \frac{\lambda M_{\text{P}}^4}{4\xi^2}
  \left(
    1+\exp\left(
      -\frac{2\chi}{\sqrt{6}M_{\text{P}}}
    \right)
  \right)^{-2}
  \;.
 \end{equation}
Inflation in models of this type is followed by harmonic oscillations 
with frequency $\omega=\sqrt{\lambda/3} M_{\text{P}}/\xi$:  
the  scalar potential is 
effectively quadratic while the amplitude of $\chi$ is large
enough, 
\begin{equation}
\label{stage}
\chi\gg M_{\text{P}}/\xi\,.
\end{equation}
 
The Universe is expanding as at the stage of matter
domination: $a\propto t^{2/3}$. In the original Higgs inflation
\cite{Bezrukov:2007ep} the
reheating happens at this stage due to decays of Higgs to the SM
particles \cite{Bezrukov:2008ut}. The interaction lagrangian between
the inflaton and any additional scalar $a$ coming from the Weyl
transformation \eqref{Weyl} takes the same form as in the Starobinsky
model of Sec.\,\ref{sec:star}: 
\be 
\label{stage1_int}
L_{\text{int}}=\frac{\chi}{\sqrt{6}M_{\text{P}}}\,\d_{\mu}a\,\d^{\mu}a~.  
\ee

If the reheating happens at this stage (Eqs. \eqref{stage1},
\eqref{stage1_int}, \eqref{stage}) 
then the decay rates of the inflaton are actually
given by eq.\,\eqref{reh}.  Here $T_{\text{reh}}$ is the temperature of the
SM plasma at the moment of equality between energy densities of
radiation and inflaton excitations. For the Higgs inflation
$T_{\text{reh}}\simeq 6\times10^{13}$\,GeV \cite{Bezrukov:2008ut}. Let us evaluate
the amount of dark radiation given the reference values of the Higgs
inflation:
\be 
\label{N_h}
\Delta N_{\text{eff}}\simeq 5.6\times 10^{-8} 
\,N \l\frac{\omega}{1.3\times 10^{-5} M_{\text{P}}}\r^3\,
\l\frac{6\times 10^{13}}{T_{\text{reh}}}\r^2.
\ee
One can observe that the axiverse with $N\sim 10^2$ is safe from the
overproduction of dark radiation in models with high enough reheating
temperature. Such models require non-gravitational couplings between
the inflaton and SM particles, like gauge and Yukawa interactions
between the inflaton (Higgs) and the SM fields in the example of Higgs
inflation. 

A side remark concerns models with the axions coupled to the SM
particles in plasma that provide with    
more efficient mechanisms of the dark radiation production.
The QCD axion with the decay constant in the
range $f_a\sim 10^{10}-10^{12}$\,GeV is a realistic example. Such axion
couples to the SM particles via dimension-5 operators suppressed by
$\Lambda=f_a$ rather than $\Lambda=M_{\text{Pl}}$. Therefore, it 
thermalizes in the SM plasma of temperature above $10^9$ GeV
which is the case for the Higgs inflation. Thus, the amount of dark
radiation in this case is defined by the axion number density in thermal
equilibrium. This leads to the value \cite{Salvio:2013iaa}
\be
\Delta N_{eff} = 0.026,
\ee
which can be measured in future CMB polarization experiments
\cite{Abazajian:2013oma,Errard:2015cxa}. At the same time, the thermal
production of string axions with $f_a\sim 10^{16}$\,GeV is still
inefficient. 

It is worth noting that the reference value of the reheating
temperature in \eqref{N_h} corresponds to the final amplitude of the inflaton
oscillations of order $\chi\sim M_{\text{P}}/\xi$ \cite{Bezrukov:2008ut}
\footnote{We do not consider values of $\xi$ much larger than those of
the Higgs inflation ($\xi \sim 10^5$) because it would lead to
strong coupling for the inflaton self-coupling, $\lambda\gtrsim
1$\,.}. 
In other words, for a quartic inflation with non-minimal coupling to
gravity, the reheating temperature cannot be lower than the reference
value in \eqref{N_h}, if the system is still at the effective 
matter domination stage provided by \eqref{stage}. Hence
eq.\,\eqref{N_h} imposes a kind of upper limit on the impact of axions
in the model with efficient reheating. 

If the reheating is less efficient than in the Higgs inflation, the
evolution comes to the second stage. There the inflaton amplitude
drops down to the value $\chi\sim M_P/\xi$ {\it before} the reheating
started, and the potential and interactions of the canonically
normalized inflaton field $\chi$ change the forms:
\be
\label{stage2}
U(\chi)=\frac{\lambda}{4} \chi^4,\quad L_{\text{int}}\sim 
\frac{\chi^2}{M_{\text{P}}^2}\d_{\mu}a\,\d^{\mu}a\,.
\ee
After this moment the axion production becomes inefficient because,
instead of the inflaton decay, we deal with the inflaton scattering
suppressed by the squared Planck mass. Moreover, the scattering rate
dilutes as $1/a^3$ due to the Universe expansion providing with
negligible overall axion production. 

Therefore at this stage, the axion production actually terminates 
providing the overall impact of axions to be of order \eqref{N_h}
calculated for the reference value of the reheating temperature
$T_{\text{reh}}\simeq 6\times10^{13}$\,GeV. One
can see that for all reasonable parameter choices we are left with a
negligible amount of axion dark radiation. Note that the real
reheating, that is a production of the SM plasma, may happen much
later, than the reference temperature indicates, but it does not
change this estimate in the slightest. Inflaton couplings to the SM
particles must be of another form than the kinetic one of
Eq.\,\eqref{stage2}, the latter is not sufficient for successful reheating.

{\bf Small non-minimal coupling $\xi\ll 1$}. Here 
the change of variables \eqref{change} may be simplified:
\be 
\frac{d\chi}{d\phi}=\frac{1}{\sqrt{1+\xi \phi^2/M_{\text{P}}^2}},
\quad \phi=\frac{M_{\text{P}}}{\sqrt{\xi}} 
\sinh{\l\frac{\sqrt{\xi}\chi}{M_{\text{P}}}\r}.
\ee
The potential \eqref{HI} transforms to
\be
\label{potential}
U(\chi)=\frac{\lambda M_{\text{P}}^4}{4\xi^2}
\tanh^4{\l\frac{\sqrt{\xi}\chi}{M_{\text{P}}}\r}.
\ee
Near the minimum, conformal factor \eqref{Weyl} can be approximated as
$\Omega^2=1+\xi \chi^2/M_{\text{P}}^2$ providing the leading interaction term
with scalars to be
\be
\label{interaction}
L_{\text{int}}=\xi \, 
\frac{\chi^2}{M_{\text{P}}^2}\d_{\mu}a\,\d^{\mu} a\,.
\ee

Potential \eqref{potential} is symmetric with respect to
$\chi\rightarrow -\chi$ so no linear terms are expected. The
production of axions in that case is inefficient due to the $M_{\text{P}}^2$
suppression of interaction\,\eqref{interaction} in accord with the
expectations we discussed right below eq.\,\eqref{Laa}.

\section{Conclusions}

In this paper, we investigate the validity of inflationary models in
the string axiverse.  Many light scalars can be produced at reheating
and later contribute to the dark radiation component of the Universe
which is strictly bounded by the recent Planck data. We found the
general conditions for the efficient production of the light scalars
at the Universe reheating. Namely, if the inflaton decays to two
axions via the dimension-5 Planck-scale suppressed operators then the
amount of the dark radiation is controlled by the reheating
temperature. For example, inflationary models with reheating via
Planck suppressed couplings of the inflaton to the SM particles (which
seems to be common in the supergravity framework) predict too much
dark radiation making them inconsistent with the cosmological
observations. We should stress that our results are directly
applicable not only in the string framework but for any light scalars
(Nambu-Goldstone bosons, dilaton) which may appear in a concrete
cosmological model.

However, there are two ways how to make the inflation consistent with
the presence of extra light scalars. The first way is to provide the
additional couplings between the inflaton and SM fields which are not
suppressed by the Planck mass. It would raise the reheating
temperature leaving no time for the axions production after
inflation. Another way around is to deal with models possessing a
symmetry which either forbids or strongly suppresses the inflaton
decay to axions (the symmetry must not prevent the
successful reheating, of course).

\vskip 0.3cm 
The authors are indebted to S.\,Dubovsky, S.\,Sibiryakov and
A.\,Starobinsky for the valuable correspondence and discussions.  


\end{document}